\documentclass[twocolumn]{aastex63}

\usepackage{lineno}
\usepackage[figuresright]{rotating}

\received{September 9, 2022}
\revised{***, ****}
\accepted{***, ****}
\submitjournal{ApJ}

\begin{document}
\title{On the Nature of the Three-part Structure of Solar Coronal Mass Ejections}

\correspondingauthor{Hongqiang Song}
\email{hqsong@sdu.edu.cn}

\author{Hongqiang Song}
\affiliation{Shandong Provincial Key Laboratory of Optical Astronomy and Solar-Terrestrial Environment, and Institute of Space Sciences, Shandong University, Weihai, Shandong 264209, China}
\affiliation{CAS Key Laboratory of Solar Activity, National Astronomical Observatories, Chinese Academy of Sciences, Beijing, 100101, China}


\author{Jie Zhang}
\affiliation{Department of Physics and Astronomy, George Mason University, Fairfax, VA 22030, USA}






\author{Leping Li}
\affiliation{CAS Key Laboratory of Solar Activity, National Astronomical Observatories, Chinese Academy of Sciences, Beijing, 100101, China}

\author{Zihao Yang}
\affiliation{School of Earth and Space Sciences, Peking University, Beijing 100871, China}

\author{Lidong Xia}
\affiliation{Shandong Provincial Key Laboratory of Optical Astronomy and Solar-Terrestrial Environment, and Institute of Space Sciences, Shandong University, Weihai, Shandong 264209, China}


\author{Ruisheng Zheng}
\affiliation{Shandong Provincial Key Laboratory of Optical Astronomy and Solar-Terrestrial Environment, and Institute of Space Sciences, Shandong University, Weihai, Shandong 264209, China}

\author{Yao Chen}
\affiliation{Institute of Frontier and Interdisciplinary Science, Shandong University, Qingdao, Shandong 266237, China}
\affiliation{Shandong Provincial Key Laboratory of Optical Astronomy and Solar-Terrestrial Environment, and Institute of Space Sciences, Shandong University, Weihai, Shandong 264209, China}








\begin{abstract}
Coronal mass ejections (CMEs) result from eruptions of magnetic flux ropes (MFRs) and can possess a three-part structure in white-light coronagraphs, including a bright front, dark cavity and bright core. In the traditional opinion, the bright front forms due to the plasma pileup along the MFR border, the cavity represents the cross section of the MFR, and the bright core corresponds to the erupted prominence. However, this explanation on the nature of the three-part structure is being challenged. In this paper, we report an intriguing event occurred on 2014 June 14 that was recorded by multiple space- and ground-based instruments seamlessly, clearly showing that the CME front originates from the plasma pileup along the magnetic arcades overlying the MFR, and the core corresponds to a hot-channel MFR. Thus the dark cavity is not an MFR, instead it is a low-density zone between the CME front and a trailing MFR. These observations are consistent with a new explanation on the CME structure. If the new explanation is correct, most (if not all) CMEs should exhibit the three-part appearance in their early eruption stage. To examine this prediction, we make a survey study of all CMEs in 2011 and find that all limb events have the three-part feature in the low corona, regardless of their appearances in the high corona. Our studies suggest that the three-part structure is the intrinsic structure of CMEs, which has fundamental importance for understanding CMEs.
\end{abstract}



\section{Introduction}
Since the very first observation of solar coronal mass ejections (CMEs) on 1971 December 14 \citep{tousey73}, great progress has been made in various aspects of CMEs, including their initiation mechanisms, acceleration processes, propagations and evolutions, as well as their space weather effects \citep{forbes06,chenpengfei11,manchester17,wyper17}. Generally, solar physics community has reached the consensus that CMEs result from eruptions of magnetic flux ropes (MFRs), which can be trigged and accelerated by ideal magnetohydrodynamic instabilities and resistive magnetic reconnections \citep{song13,song15a,song18a,chengxin20}, while some issues remain elusive due to their complexity and limited observations. Researchers had thought that it is straightforward to understand the morphological structures of CMEs in the white-light coronagraph, as the brightness of K-corona originates from Thomson scattering of free \textbf{electrons} and depends on the electron density around the plane of sky near the Sun \citep{hayes01}. The bright front, dark cavity, and bright core have been explained as the pileup of ambient plasma, MFR, and prominence, respectively.

However, direct observations of CME front formation remain rare, and many CMEs with the three-part structure are unrelated with prominence eruptions \citep{howard17}. Recent studies employing dual-viewpoint observations proposed a unified explanation on the nature of the three parts of CMEs \citep{song19a,song19b,song22b}. The explanation is tightly correlated with the MFRs and overlying magnetic arcades, which play key roles for CME initiation \citep{amari18,jiangchaowei21}. MFRs are in the equilibrium between upward lifting and downward pulling Lorentz forces prior to eruption. The lifting force is provided by the azimuthal current inside the MFR and its image current below the photosphere, and the pulling force is produced by the overlying arcades with potential field \citep{chenyao06b}. When MFRs start to rise, the overlying arcades are stretched, piling up the surrounding plasmas and forming the CME front \citep{forbes00,chenpengfei09}. The MFR is observed as the CME core, and the zone between the MFR and front corresponds to the dark cavity \citep{song19a,song19b,song22b}. The MFR can expand and occupy the cavity space, in the meantime, the cavity magnetic field can be transformed into the outer shell of MFR through magnetic reconnections in the current sheet underneath CMEs \citep{linjun00}. Thus the relative volume of dark cavity will decrease gradually and disappear eventually \citep{song17b}.

The above new explanation answers why CMEs can exhibit the three-part structure without prominences and how they lose the dark cavity feature, but lacks complete evidence. Besides, it also suggests that each CME should possess the three-part feature in the early eruption stage, i.e., the three-part structure is the intrinsic structure of CMEs. Here, we provide strong evidence for the explanation through an intriguing CME that was recorded by multiple space- and ground-based instruments, including the Atmospheric Imaging Assembly (AIA) \citep{lemen12} on board the Solar Dynamics Observatory (SDO) \citep{pesnell12}, the Extreme Ultraviolet Imager (EUVI) \citep{howard08} on board the Solar Terrestrial Relations Observatory (STEREO) \citep{kaiser05}, the Large Angle and Spectrometric Coronagraph (LASCO) \citep{brueckner95} on board the Solar and Heliospheric Observatory (SOHO) \citep{domingo95}, as well as the coronal solar magnetism observatory K-coronagraph (K-Cor) and Coronal Multichannel Polarimeter (CoMP) \citep{tomczyk08} at the Mauna Loa Solar Observatory (MLSO). SDO, SOHO and MLSO observe the Sun from the Earth perspective, while STEREO from different directions. A preliminary survey study is also conducted to examine the prediction of our explanation.

\section{Instruments}
The AIA onboard the SDO takes images of the Sun with an FOV of 1.3 $R_\odot$ through seven EUV channels, including 131 \AA\ (Fe XXI, $\sim$10 MK), 94 \AA\ (Fe XVIII, $\sim$6.4 MK), 335 \AA\ (Fe XVI, $\sim$2.5 MK), 211 \AA\ (Fe XIV, $\sim$2.0 MK), 193 \AA\ (Fe XII, $\sim$1.6 MK), 171 \AA\ (Fe IX, $\sim$0.6 MK), and 304 \AA\ (He II, $\sim$0.05 MK). These AIA images have a spatial resolution of 1.2$\arcsec$ and a temporal cadence of 12 s. The 131 \AA, 171 \AA, 304 \AA, and 193 \AA\ images are used here and they are processed to level 1.5 through standard codes in the solar software. The solar software package is available online\footnote{http://www.lmsal.com/solarsoft/}.

The K-Cor records the coronal polarization brightness (pB) formed by Thomson scattering of photospheric light from free electrons. It covers an FOV of 1.05--3 $R_\odot$ in the passband of 7200--7500 \AA\ with 5.5$\arcsec$ pixel size and a nominal cadence of 15 s. The CoMP can perform the spectropolarimetric observations at infrared wavelengths (10747 and 10798 \AA). Its FOV is 1.05--1.35 $R_\odot$ with a 4.35$\arcsec$ pixel size. The Doppler-velocity map is obtained through an analytical Gaussian fitting to the three-point intensity profile at each pixel \citep{tianhui13}. The flare X-ray data are from the Geostationary Operational Environment Satellite (GOES) that provides the integrated full-disk soft X-ray emission from the Sun.


\section{Observations and Results}
The solar eruption of interest occurred on 2014 June 14 when STEREO-A (B) was $\sim$161$^{\circ}$ (164$^{\circ}$) west (east) of the Earth as shown in Figure 1. The eruption originated from NOAA Active Region 12094 located at the heliographic coordinates $\sim$S13W74 when observed from the STEREO-B perspective as displayed in the EUVI 195 \AA\ image (top-left panel). Note that the white-dashed line denotes the solar limb observed from the Earth direction. It is an exact limb event ($\sim$S13E90) from the Earth as confirmed by the AIA 193 \AA\ image (bottom-left panel). The eruption produced an M1.4 class soft X-ray flare on GOES scale (top-right panel) and a CME with a linear speed of 732 km s$^{-1}$ in the field of view (FOV) of LASCO (Coordinated data analysis workshops, CDAW\footnote{https://cdaw.gsfc.nasa.gov}). The red-dashed line delineates the CME front, and this CME does not exhibit the typical three-part structure in the C2 image (bottom-right panel).

The initial eruption process is recorded well by the AIA, CoMP, and K-Cor as shown in Figure 2. The top-left panel presents the 171 \AA\ image at 19:23:59 UT, displaying several bunches of coronal loops prior to the eruption. We pay our attention on the central loops as pointed with two arrows, which have clear correlation with the formation of CME front. The CoMP Doppler-velocity image is also presented in this panel, with its FOV delineated by the white rectangle. The coronal loops started to rise from $\sim$19:25 UT onward, and the snapshot at 19:27:59 UT is presented in the top-middle panel. The leading front of the loops is delineated with purple-dotted line. The CME is associated with obvious Doppler blue shift as displayed in the CoMP image. Please note that the Doppler shift values do not represent the line-of-sight component of the CME propagation speed, and their changes should be largely associated with the coronal response to the lateral expansion and mass eruptions of CMEs \citep{tianhui13}. Top-right panel displays the K-Cor base-difference image at 19:27:58 UT (with 19:23:10 UT as the base), when only the CME front can be observed. The purple-dotted line in the top-middle panel is overlaid on the K-Cor image, corresponding to the inner edge of the CME front. The outer edge is delineated with blue-dotted line that is also overlaid on the 171 \AA\ image in the top-middle panel, which locates higher than the coronal loops. These observations clearly show that the CME front originates from the pileup of plasmas surrounding the rising loops, rather than the MFR that appears later.

The bottom-left panel presents the AIA 131 \AA\ image at 19:27:56 UT, also with the blue and purple dotted lines overlaid. The hot-channel MFR starts to appear from $\sim$19:28:20 UT onward, and becomes very obvious at 19:33:32 UT as delineated with red-dotted line in the bottom-middle panel. The accompanying animation (the left part in each frame) displays the eruption process with the composite observations of AIA 131 \AA , 171 \AA , and 193 \AA. The K-Cor image at the same time is displayed in the bottom-right panel with the red-dotted line overlaid, which unambiguously demonstrates that the hot-channel MFR is observed as the CME core at this time. The CME front is also identifiable as marked with blue-dotted line. Therefore, this CME exhibits the typical three-part morphological structure in the early eruption stage. Its dark cavity disappeared gradually when propagating outward, which can be examined continuously through the accompanying animation (the right part in each frame). That is why the CME does not have the typical three-part feature in the C2 image as shown in Figure 1.

Our observations provide clear connections between the CME front formation and the coronal loop rising, direct correspondence between the CME core and hot-channel MFR, as well as the disappearance of CME cavity through combining observations of both low and high corona. This supports that the CME cavity is a low-density zone between the MFR and overlying coronal loops in the early eruption stage. Our new explanation successfully answers why CMEs can possess the three-part structure no matter whether prominences are involved or not. Both situations can be described with a schematic drawing as shown in Figure 3, where the top and bottom panels display the cases associated without and with a prominence, respectively.

The top-left panel shows the configuration prior to eruption when an MFR (filled with blue) is in equilibrium between upward lifting and downward pulling Lorentz forces. The thick line represents the outer most coronal loops with potential field and the thin lines represent the sheared magnetic field of the low-density zone, as well as the twisted MFR. The loops start to rise after eruption onset, and the surrounding plasmas are piled up, forming the CME front as denoted with brown in the other top panels. The low-density zone exists in the early eruption stage, thus the CME can be recorded with the classical three-part feature in the white-light images, as presented in the top-middle panel. As mentioned, the field lines surrounding the MFR can be transformed into the outer part of the MFR through magnetic reconnections in the current sheet (depicted with the vertical dotted line) connecting the MFR and flare loops, thus the MFR grows and expands during the eruption. The MFR can occupy the low-density zone eventually as shown in the top-right panel, resulting in the disappearance of dark cavity \citep{song17b}.

The bottom two panels display the corresponding situations with a prominence (an ellipse filled with red) supported by the MFR. In this case, both the prominence and MFR can be recorded as the CME cores in the early eruption stage. The prominence could appear as a sharp core, while the MFR, a fuzzy core, due to their density characteristics \citep{song19a}. Similarly, the MFR rises and stretches its overlying loops, forming the CME front, and the original low-density zone can disappear due to the MFR growth and expansion. In the meantime, the prominence material can drain back to the solar surface \citep{ouyang17} and/or its density can decrease obviously due to expansion during propagation outward. These lead to the disappearances of the dark cavity and sharp core, i.e., the CME loses the three-part feature eventually as pointed with the red arrow. The light red represents a prominence with relatively lower densities. However, the prominence can also maintain obviously higher densities compared to the MFR due to few draining back of its plasma and/or less expansion. In this case, the CME could keep the three-part feature in the high corona after the original low-density zone is transformed into the MFR shell. This situation is illustrated in the panel pointed with the green arrow. The deep red represents a prominence with relatively higher densities. At this time, the prominence and MFR are observed as the bright core and dark cavity, respectively, i.e., the three-part feature can be explained with the traditional opinion. Therefore, our explanation can answer why prominence-related CMEs do or do not exhibit the three-part structure in the high corona.

According to the above explanation, we believe that probably all CMEs should exhibit the three-part feature in the low corona, except for the stealth CMEs that do not have the distinct low coronal signatures \citep{masuli10} and narrow CMEs that might be just jet-like structure. Usually it is difficult to distinguish the three parts for disk events due to the projection effect, then we expect that all limb CMEs possess the three-part appearance in the low corona. To examine this prediction, we conduct a survey study based on CMEs recorded by LASCO in 2011. In total 1990 CMEs are identified manually in this year (CDAW), while most events are poor or very poor ones and unsuitable for the morphological analysis. After removing these events we inspect the remaining 535 CMEs and find 98 (18.3\%) of them exhibiting the typical three-part feature in C2 images, a similar fraction to previous study by \cite{vourlidas13}. The 437 non-three-part CMEs include 118 narrow events with angular width no more than 40$^{\circ}$ and 319 normal events. After inspecting the normal events one by one, we find 28 of them were located at the solar limb unambiguously. All of these events possess the three-part feature in the composite images of AIA 304 \AA\ or 131 \AA\ and 171 \AA\ (or 193 \AA) when they are associated with eruptions of prominences or hot channels \citep{zhangjie12}. The event No., CME first appearance time in the C2 image (CDAW), source region (AR for active region and QS for quiet-Sun region), as well as the ejecta (P for prominence and HC for hot channel) are listed sequentially in Table 1 for the 28 events.

\begin{table}[!htbp]
\begin{minipage}[t]{\columnwidth}
\caption{The information of 28 limb CMEs without typical three-part structure in the C2 images but with the three-part feature in the EUV images}.
\tabcolsep=3pt
\begin{tabular}{cccc}
  \hline \hline
  No. & First in C2 & Source & Ejecta\\
  \hline
  01 & Jan 28 01:25 UT & AR & P\\
  02 & Feb 11 22:12 UT & AR & P\\
  03 & Feb 16 23:11 UT & AR & HC\\
  04 & Mar 04 05:36 UT & AR & P\\
  05 & Mar 07 20:00 UT & AR & P\\
  06 & Mar 08 04:12 UT & AR & HC\\
  07 & Mar 27 05:36 UT & AR & P\\
  08 & Apr 07 12:00 UT & AR & HC\\
  09 & Apr 09 18:00 UT & AR & HC\\
  10 & May 03 16:51 UT & QS & P\\
  11 & May 09 20:57 UT & AR & HC\\
  12 & May 11 02:48 UT & QS & P\\
  13 & May 18 18:24 UT & AR & P\\
  14 & May 29 21:24 UT & AR & HC\\
  15 & Jun 05 16:59 UT & QS & P\\
  16 & Jun 07 06:49 UT & AR & P\\
  17 & Jun 12 14:48 UT & AR & P\\
  18 & Jul 09 06:12 UT & AR & P\\
  19 & Jul 09 16:36 UT & AR & P\\
  20 & Jul 10 12:00 UT & QS & P\\
  21 & Aug 06 10:24 UT & QS & P\\
  22 & Aug 17 02:48 UT & AR & P\\
  23 & Sep 22 10:48 UT & AR & HC\\
  24 & Oct 18 08:12 UT & AR & P\\
  25 & Nov 07 04:12 UT & AR & HC\\
  26 & Dec 04 20:12 UT & AR & P\\
  27 & Dec 08 09:24 UT & QS & P\\
  28 & Dec 30 20:57 UT & AR & HC\\
  \hline
  \hline
\end{tabular}
\end{minipage}
\end{table}

In Table 1, 19 events are associated with prominence eruptions, including 13 (6) events from active (quiet-Sun) regions, and 9 events result from hot channel eruptions, which all originate from active regions. As the event displayed in Figures 1 and 2 is associated with a hot channel eruption, here we select a prominence eruption (Event 7) from Table 1 as the representative event and present it in Figure 4. The prominence eruption occurred on March 27, and generated a CME without typical three-part structure in LASCO image as presented in the left panel. The CME front is delineated with the red-dashed line. The right panel displays the composite image of two passbands, which exhibits the three-part feature with 304 \AA\ and 193 \AA\ showing the CME core (prominence) and front, respectively. Note that the base-difference image is used for 193 \AA\ to show the front clearly. The accompanying animation, created with the JHelioviewer software \citep{muller17}, displays the whole eruption process from 04:35 to 05:35 UT with composite observations of AIA 304 \AA\ and 193 \AA.

\section{Summary and Discussion}
We report an intriguing solar eruption occurred on 2014 June 14 that was recorded by multiple space- and ground-based instruments seamlessly. Our observations clearly demonstrate that the pileup of plasmas surrounding the coronal loops results in the formation of CME front, and unambiguously present the correspondence between the hot-channel MFR and CME core in white-light images. Naturally, the low-density zone between the coronal loops and hot channel is observed as the dark cavity of the CME. These perfectly answer the nature of three-part structure of CMEs, and suggest that this structure is their intrinsic structure and each normal CME should has the three-part feature. Our survey study shows that each limb CME (except for stealth CMEs and narrow ones) has the three-part feature in the early eruption stage, agreeing with our prediction. The growth and expansion of hot-channel MFR reduce the relative volume of the low-density zone \citep{song17b}. This explains the disappearance of dark cavity and answers why not all CMEs possess the three-part feature in the high corona.

It is necessary to discuss why a low-density zone exists between the CME front and MFR. A reverse current model has been reported to explain the CME cavity formation \citep{haw18}, which proposed that the cavity forms because the rising electric current in the MFR induces an oppositely directed electric current in the background plasma. The magnetic force between the two electric currents propels the background plasma away from the MFR, creating the low-density zone, i.e., the CME dark cavity. This model is supported by laboratory experiments, three-dimensional numerical simulations, and an analytic model \citep{haw18}. According to this model, the CME cavity could disappear due to the decrease of electric current intensity along with time. More detailed observations on the disappearance process of the dark cavity are necessary to clarify its nature.

China successfully launched the Advanced Space-based Solar Observatory (ASO-S) \citep{ganweiqun19b} on 2022 October 9. The ASO-S, nicknamed Kuafu-1, is equipped with three payloads: the Full-disk vector MagnetoGraph (FMG) \citep{dengyuanyong19}, the Lyman-$\alpha$ Solar Telescope (LST) \citep{fengli19}, and the Hard X-ray Imager (HXI) \citep{suyang19}. The LST consists of three instruments: the Solar Disk Imager (SDI), Solar Corona Imager (SCI) and White-light Solar Telescope (WST), which cover an FOV of 0--2.5 $R_\odot$ and can seamlessly observe CMEs in their early eruption stage through both white-light and Lyman-$\alpha$ passbands. ASO-S provides us an excellent opportunity to continue the study of three-part structure of CMEs in solar cycle 25.


\acknowledgments We are grateful to the anonymous referee for the comments and suggestions that helped to improve the original manuscript significantly. We thank Profs. Pengfei Chen and Xin Cheng at Nanjing University for their helpful discussions. This work is supported by the NSFC grants U2031109, 11790303 (11790300), and 12073042. H.Q.S is also supported by the CAS grants XDA-17040507 and the open research program of the CAS Key Laboratory of Solar Activity KLSA202107. Data from observations are courtesy of SDO, MLSO, SOHO, STEREO, and GOES.





\begin{figure*}[htb!]
\epsscale{0.9} \plotone{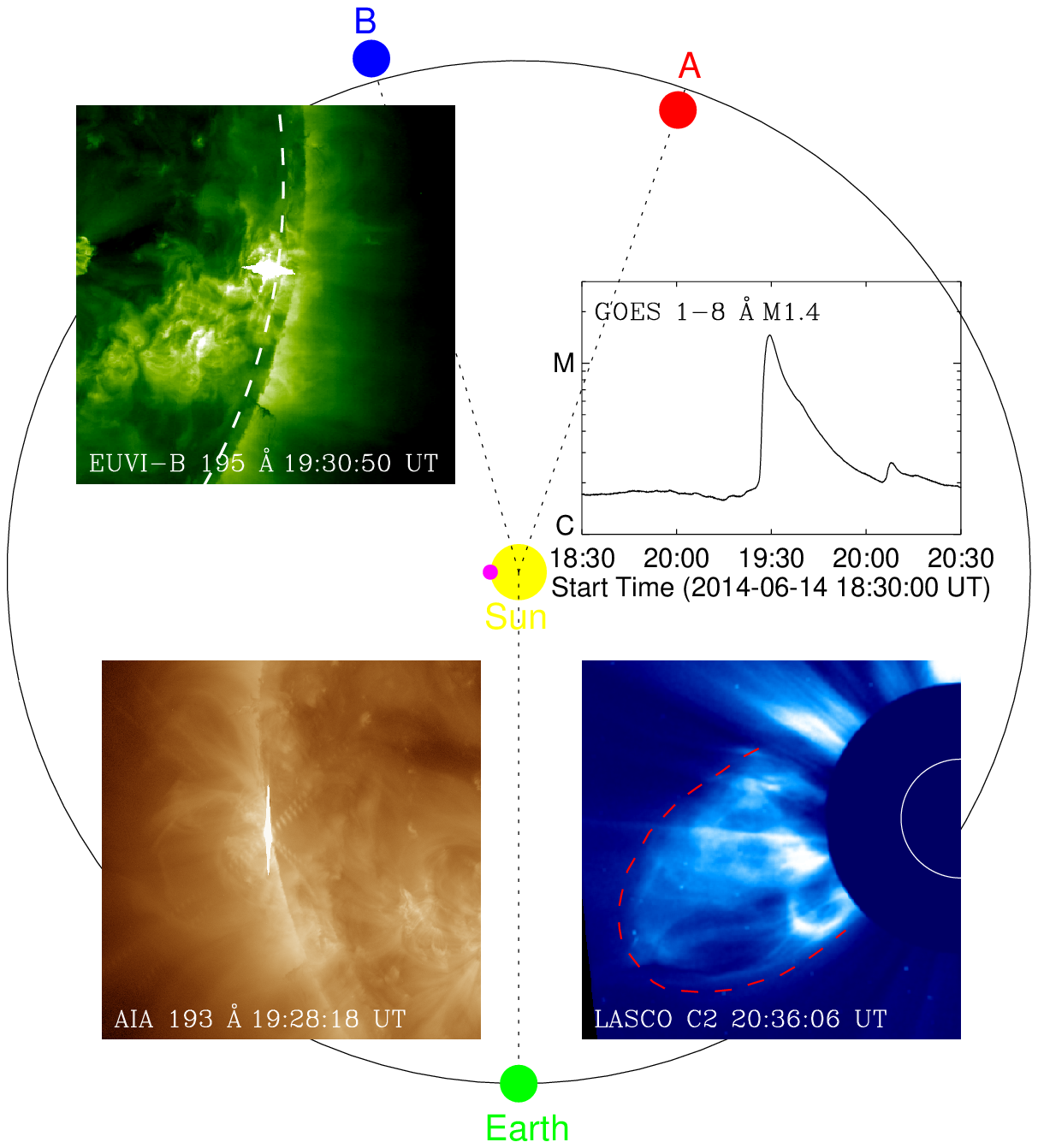} \caption{Positions of the Earth (SDO and SOHO) and STEREO in the ecliptic plane on 2014 June 14 and their observations. (Top Left) EUVI-B 195 \AA\ image, showing the source location of solar eruption. (Bottom Left) AIA 193 \AA\ image, confirming a limb event from the Earth perspective. (Top right) GOES soft X-ray 1-8 \AA\ flux, displaying an M1.4 class flare. (Bottom right) LASCO C2 image, presenting a CME without the typical three-part structure. \label{Figure 1}}
\end{figure*}

\begin{figure*}[htb!]
\epsscale{1.0} \plotone{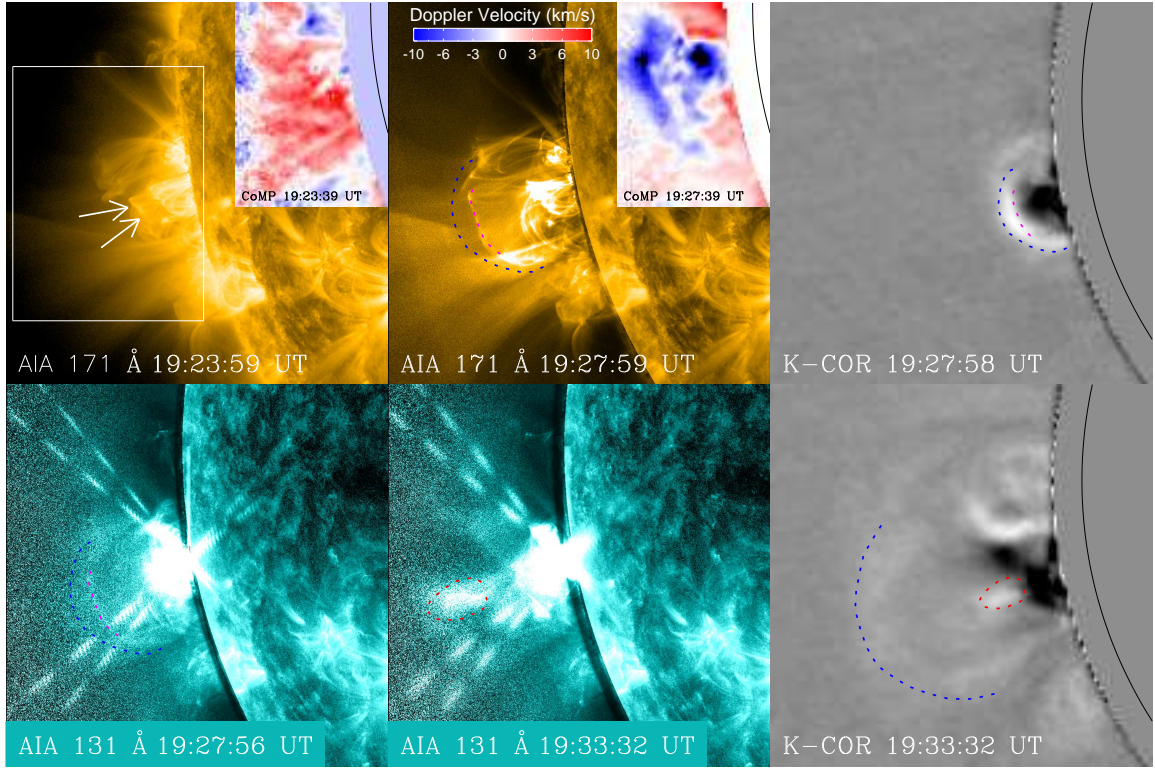} \caption{Observations of the CME prior to and during eruption. (Top left) AIA 171 \AA\ image showing the coronal loops prior to eruption at 19:23:39 UT, with the inner panel displaying the simultaneous CoMP observation of Doppler-velocity. (Top middle) AIA 171 \AA\ image showing the rising loops during eruption at 19:27:59 UT, also along with the CoMP Doppler-velocity image. (Top right) Base-difference image of K-Cor at 19:27:58 UT, displaying the spatial relations of CME front and coronal loops in top-middle panel. (Bottom left) AIA 131 \AA\ image at 19:27:56 UT. (Bottom middle) AIA 131 \AA\ image at 19:33:32 UT with red-dotted line delineating the hot-channel MFR. (Bottom right) Base-difference image of K-Cor at 19:33:32 UT, showing the typical three-part structure of the CME. The left part of each frame of the animation displays the complete eruption process from 19:22 to 19:39 UT with the composite observations of AIA 131 \AA, 171 \AA, and 193 \AA, while the right part of each frame presents the evolution process of the CME from 19:23 to 19:47 UT through the base-difference images of K-Cor. The duration of the animation is 8 s. \label{Figure 2}}
\end{figure*}

\begin{figure*}[htb!]
\epsscale{.9} \plotone{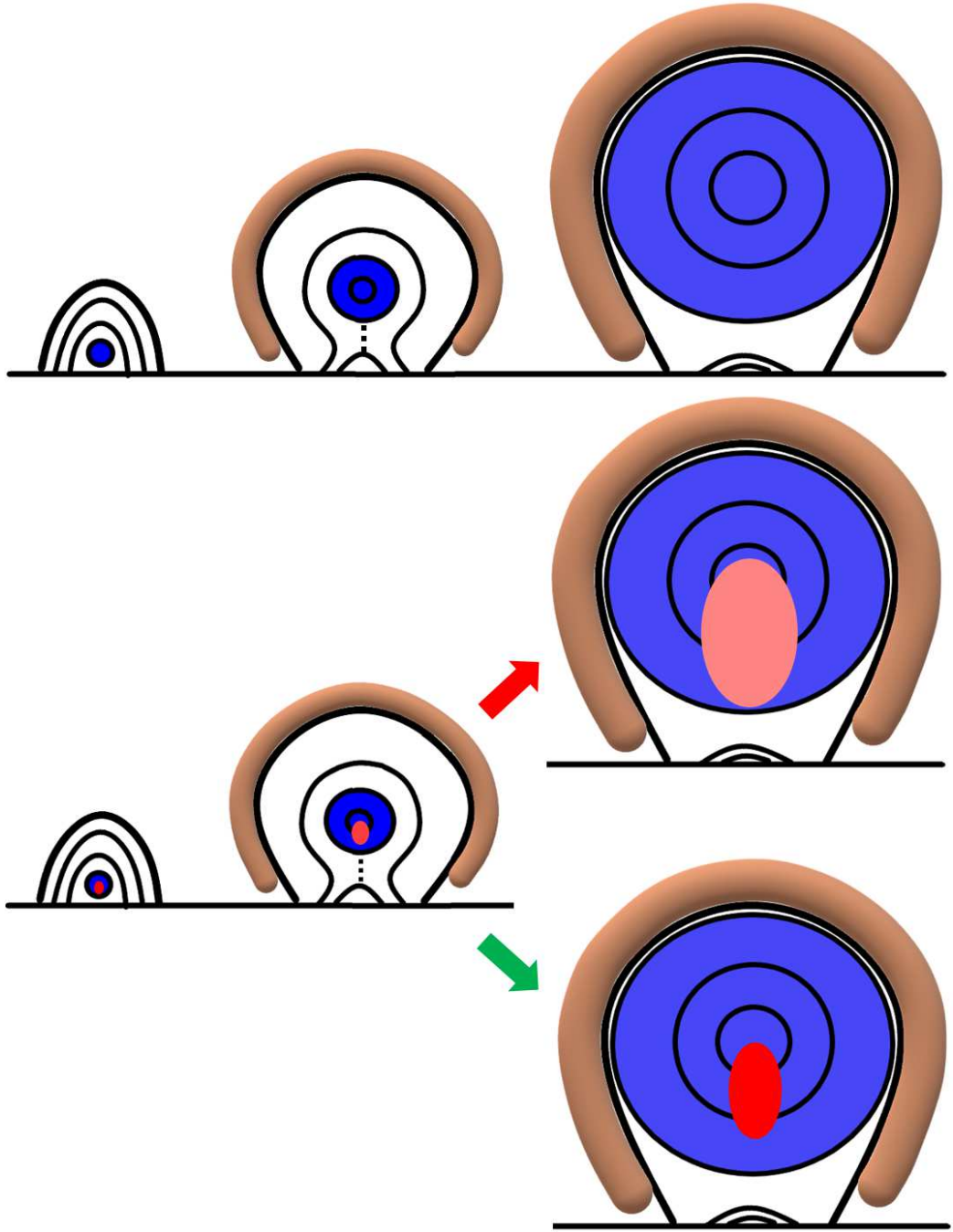} \caption{A 2-dimensional schematic sketch for the intrinsic structure of CMEs. The top panels display the case without a prominence. (Left) The MFR (circle filled with blue) is in equilibrium prior to eruption. (Middle) The MFR , corresponding to the CME core, rises and stretches the overlying coronal loops, and the background plasmas are piled up surrounding the loops, corresponding to the CME front (brown). The vertical dotted line represents the current sheet. (Right) The low-density zone is occupied by MFR expansion and magnetic reconnections in the current sheet, leading to the disappearance of the CME cavity. The bottom panels display the case with a prominence as represented by the ellipse filled with red. (Left) The MFR and prominence are in equilibrium prior to eruption. (Middle) The MFR and prominence, corresponding to the CME core, rise and stretch the overlying coronal loops, and the background plasmas are piled up surrounding the loops, corresponding to the CME front. The vertical dotted line represents the current sheet. (Right) The low-density zone is occupied by MFR expansion and magnetic reconnections in the current sheet. In some cases as pointed with the red arrow, the prominence density decreases obviously due to the expansion and material draining back to solar surface, thus the CME loses the three-part feature in the high corona. In other cases as pointed with the green arrow, the prominence keeps relatively higher density, and the CME could still exhibit the three-part feature with the prominence and MFR being observed as the bright core and dark cavity, respectively. \label{Figure 3}}
\end{figure*}

\begin{figure*}[htb!]
\epsscale{0.9} \plotone{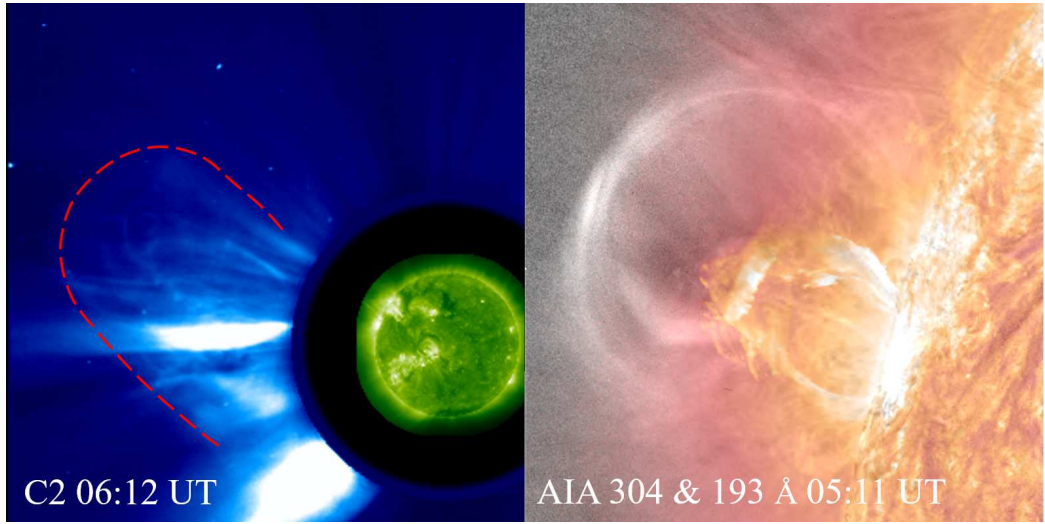} \caption{A representative event selected from Table 1 (Event 7 occurred on 2011 March 27). (Left) The LASCO C2 image, showing a CME without typical three-part structure in the white-light image. (Right) The composite image of AIA 304 \AA\ and 193 \AA, displaying the three-part feature in the EUV passbands. The animation displays the complete eruption process from 04:35 to 05:35 UT with the composite observations of AIA 304 \AA\ and 193 \AA. The duration of the animation is 3 s, which is created with the JHelioviewer software \citep{muller17}. \label{Figure 4}}
\end{figure*}

\end{document}